\documentclass[journal=jpccck,manuscript=article,layout=twocolumn,spacing=double]{achemso}
\usepackage{setspace} 
\usepackage[T1]{fontenc}
\usepackage[version=4]{mhchem}
\usepackage{amsmath,amssymb,bm}
\usepackage{mathtools}
\usepackage{graphicx}
\usepackage{dcolumn}
\usepackage{color}
\usepackage{hyperref}

\newif\ifgraph
\graphtrue
\SectionNumbersOn
\let\oldmaketitle\maketitle
\let\maketitle\relax
\author{Poulami Bag}
\affiliation{
Department of Chemistry,
Presidency University,
Kolkata 700073, India
}
\email{poulamib.rs@presiuniv.ac.in}
\date{\today}
\title{Effect of Convection Rolls in Motility-Induced Phase Separation of Active Janus Particles}
\begin{document}
\begin{singlespace}
\twocolumn[
\begin{@twocolumnfalse}
\oldmaketitle
\begin{abstract}
We numerically study motility-induced phase separation of active particles in two-dimensional convection rolls. We analyse local packing-fraction distributions,density fluctuations, the corresponding phase diagrams, and diffusivity curves to characterise the interplay between self-propulsion, global packing fraction, and advection strength. In the weak-flow regime, the system exhibits phase separation characterised by bimodal density distributions, slowly decaying density fluctuations, and a sharp reduction in diffusivity. Increasing advection suppresses clustering by enhancing particle transport and reducing trapping, leading to a shift in the critical self-propulsion velocity for motility-induced phase separation and a shrinkage of the spinodal region. Beyond the intuitive suppression of clustering by weak-flow advection, our results reveal several non-trivial phenomena including a reentrant phase behaviour where extremely high self-propulsion hinders motility-induced phase separation by facilitating particle escape from dense regions. We also observe that the density distributions strongly depend on roll periodicity. These findings demonstrate that convection rolls provide an effective means to control non-equilibrium collective behaviour in active matter. 

\end{abstract}
\end{@twocolumnfalse}
]
\section{Introduction} \label{intro}
Active particles at the nano- and micro-scales have drawn considerable interest for their complex out-of-equilibrium dynamics and possible relevance to soft matter physics, biotechnology, and nanotechnology \cite{Wang, Marchetti, Bechinger, photo-system-1, photo-system-2}. Examples range from biological micro-swimmers, bacteria, algae, and sperm cells to synthetic systems such as catalytic Janus colloids \cite{nanoscale, Granick, Muller, Paxton, Sen, Gompper, misko-1, misko-2, misko-3, ap1,Popov,Baran,Bo}. What sets these particles apart from passive matter is their ability to harvest energy from the surrounding environment and transform it into sustained, directed motion. This continuous energy input pushes them far from thermodynamic equilibrium and gives rise to rich dynamical behaviors \cite{ratchet, Ai-rat, Reichhardt1, JPCL, P-Bag, MSshort, our-review}.

The collective phenomena that emerge in active matter are remarkably diverse, such as rectified transport \cite{MSshort,our-review,ratchet,Ai-rat, P-Bag} through asymmetric geometries, chemotaxis-like  \cite{ stimuli-1, stimuli-2, stimuli-3} migration driven by density-dependent motility and spontaneous self-organisation into dynamic patterns. Among these, motility-induced phase separation (MIPS) has emerged as an intriguing characteristic that demonstrates of how activity alone can produce large-scale order. \cite{Lowben-e1, Nayak_mips, Castro1, Castro2, singlefile, QS}. In MIPS, particles that interact only through steric repulsion separate into coexisting dense and dilute regions. The underlying mechanism is a positive feedback loop between local crowding and reduced motility. Collisions slow particles down, trap them in dense regions, and draw in more neighbours. Once the global packing fraction crosses a threshold value, this random accumulation triggers macroscopic phase separation without any attractive forces. A substantial body of research \cite{Nayak_mips, Castro1, Castro2, van, zhou, Yoshinaga, Ma, Musacchio, Duan, Sansa, Su, Solon, Klamser} now exists on MIPS. Nearly all of this work, however, assumes a quiescent background. Active suspensions in the real world are rarely still. Even a gentle ambient flow can advect particles, modify density fluctuations, and thereby influencing the clustering instability.

\begin{figure*}[t]
	\centering	\includegraphics[width=0.85\textwidth]{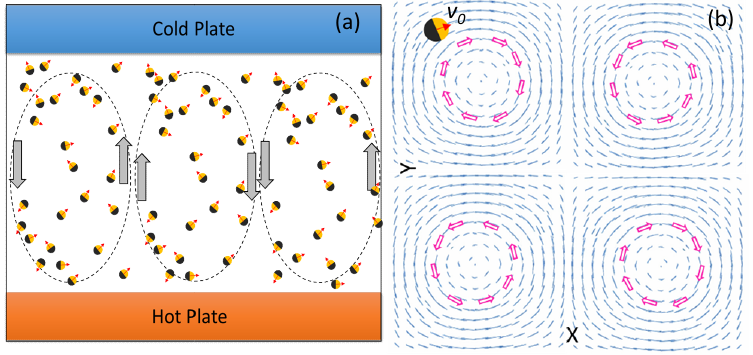}
    \caption{Schematic of the model and convective flow field. (a) Unit flow cell generated by a hot plate (bottom) and a cold plate (top), showing active Janus particles advected by the underlying periodic convective pattern. The flow is defined by the stream function $\psi(x,y)$, with velocity field $\bm{u}(\bm{r},t) = (\partial_y \psi, -\partial_x \psi)$. Dashed contours indicate streamlines forming convection rolls, while arrows denote upward (hot) and downward (cold) fluid motion. (b) Corresponding periodic array of four counter-rotating convection subcells.}
    \label{Fig1}
\end{figure*}

Previous studies provide useful insights into the behaviour of active particles in linear (e.g., Couette flow) or nonlinear (e.g., Poiseuille flow, convection rolls) flows at low Reynolds numbers \cite{zottl1, zottl2, shear, ghosh1, t-deb, Asheichyk, Asheichyk2}. Most of them focus on non-interacting active particles, and a few consider collective motion. Considering all the advances, \cite{ Sarracino,Torney,Lauga,young,solomon,yli-1,Y-Li,Q-yin} here we examine how the convection rolls reshape MIPS. The flow introduces a competing transport where particles are carried along by the solvent as well as propelled by their own activity.  Particles are also subjected to a flow- induced advection torque $(\Omega_{\text{ad}})$, that impacts particles' orientational dynamics. A natural dimensionless measure of this competition is $\chi = u_0/v_0$, the ratio of the advection velocity $(u_0)$ to the self-propulsion velocity $(v_0)$. Depending on the value of $\chi$, advection can suppress clustering, shift the phase boundary, or change the morphology of the separated domains altogether.

Our goal is a systematic investigation of how MIPS stability gets impacted by activity $(v_0)$, mean packing fraction $(\phi)$, and advection strength $(u_0)$. By systematically mapping local packing-fraction distributions and density fluctuations, effective diffusivities, and complete phase diagrams, we demonstrate that convective flow serves as  robust external control parameter.  \\

The outline of this paper is as follows. In Sec.2, we present the model describing the interacting dynamics of active Janus particles inside convection rolls. In Sec.3, we present and analyse the numerical results, focusing on the characterisation of MIPS. Finally, we summarise our findings in Sec.4.

\section{Model}
 We consider a two-dimensional suspension of $N$ identical active Janus particles of fixed radius $r_0$ confined to a square domain of linear size $L_{box}$ with periodic boundary conditions. The system represents a minimal active particle model perturbed by an incompressible convective flow field. Hydrodynamic interactions between particles are neglected, and inertia is assumed to be negligible, placing the dynamics in the overdamped regime appropriate for active microswimmers.

Each particle self-propels at a constant speed $v_0$ along its instantaneous body orientation $\bm{e}(\theta_i) = (\cos\theta_i,\sin\theta_i)$. The coupled translational and rotational dynamics of the $i$th particle are governed by the overdamped Langevin equations,
\begin{equation}
\dot{\bm{r}}_i = \bm{u}(\bm{r}_i,t) + \sum_{j \ne i} \bm{F}_{ij} + v_0 \bm{e}(\theta_i) + \sqrt{2D_t}\,\bm{\xi}_i(t),
\label{eq:position}
\end{equation}
\begin{equation}
\dot{\theta}_i = \Omega_{ad} + \sqrt{2D_r}\,\xi_i^{\theta}(t),
\label{eq:orientation}
\end{equation}
%
The first term in Eq.~(\ref{eq:position}), $\bm{u}(\bm{r}_i,t)$, describes the advection velocity. The second term represents short-range steric repulsion between particles. Interactions are modeled using the Weeks-Chandler-Anderson (WCA) potential,
\begin{equation}
V_{\mathrm{WCA}}(r) = 
\begin{cases} 
4\varepsilon \left[ \left(\frac{\sigma}{r}\right)^{12} - \left(\frac{\sigma}{r}\right)^6 \right] + \varepsilon, & r \le 2^{1/6}\sigma \\ 
0, & r > 2^{1/6}\sigma 
\end{cases}
\label{eq:WCA}
\end{equation}
where $r_{ij}=|\bm{r}_i-\bm{r}_j|$ is the interparticle distance, $\varepsilon$ sets the interaction strength, and $\sigma = 2r_0$ is the particle diameter. The repulsive force between a pair of particles is, $\bm{F}_{ij} = - \nabla_i V_{\mathrm{WCA}}(r_{ij})$
The WCA interaction accounts for excluded-volume effects without
attractive interactions, thereby enabling MIPS to arise purely from activity and steric effects.

The translational and rotational fluctuations, $\bm{\xi}_i(t)$ and $\xi_i^{\theta}(t)$, are independent Gaussian white noises satisfying:
$$
\langle \bm{\xi}_i(t) \rangle = \bm{0}, \quad \text{and} \quad
\langle \xi_{i\alpha}(t)\xi_{j\beta}(t') \rangle = \delta_{ij}\delta_{\alpha\beta}\delta(t - t')
$$
 
with $\alpha,\beta = x,y,\theta$. The parameters $D_t$ and $D_r$ denote the translational and rotational diffusion coefficients, respectively.

The persistence time of active motion is given by $\tau_\theta = 1/D_r$, leading to the persistence length $l_\theta=v_0 \tau_\theta$, which determines the crossover between the ballistic and diffusive regimes of single-particle motion.

To model convection rolls, we consider a two-dimensional incompressible flow field constructed from the stream function $\psi(x,y)$ as,
\begin{equation}
\psi(x,y) = \frac{u_0 L}{2\pi}\sin\!\left(\frac{2\pi x}{L}\right) \sin\!\left(\frac{2\pi y}{L}\right),
\end{equation}
\begin{equation}
\bm{u}(\bm{r},t) = \left(\partial_y \psi,-\partial_x \psi\right).
\end{equation}
Here, $L$ represents the period of the convection rolls and varying $u_0$ systematically tunes the advection strength. The characteristic roll vorticity is $ \Omega_{ad}=2\pi u_0/L$, corresponding to the maximum magnitude of the local angular velocity field. $D_L$ is the advective diffusivity defined as, $D_L = u_0 L/2\pi$. \\
This construction guarantees $\nabla \cdot \bm{u} = 0$. The resulting velocity field consists of a periodic array of counter-rotating convection rolls resembling Rayleigh–Bénard cells. 

The global packing fraction is given by,
$\phi=N \pi r_0^2/L_{box}^2$. The $l_\theta$ value must be compared with the other characteristic length scales, including interparticle separation $l_d$  and the mean free path between a pair of collisions $l_c$, which are derived from standard kinetic theory principles. Assuming that each particle occupies a square of side length $l_d$ inside the simulation box, therefore $l_d = \sqrt{L_{box}^2/N}$. Here, $l_d \propto \rho^{-1/2}$, (where $\rho=N/L_{box}^2$ is the number density) represents the average spacing scale between particles in a 2D system, and $l_c =1/(\rho\sigma) = L_{box}^2/(N\sigma)$. The particle diameter is taken small enough to prevent excluded-volume effects from controlling the dynamics of the system. The characteristic lengths can be expressed in terms of the average packing fraction ${\phi}$ as, $l_d = \sqrt{\pi r_0^2/\phi}$ and $l_c = \pi r_0/2\phi$.  
In this representation, both the average interparticle separation and the mean free path depend only on the packing fraction. Consequently, for a fixed value of $\phi$, the relative balance between persistence, interparticle spacing, and collisional effects remains unchanged. This implies that varying the total number of particles $N$ does not influence the dimensionless ratios $l_d / l_\theta$ and $l_c / l_\theta$, and therefore does not modify the underlying dynamical behavior of the system.

We limit our study to suspension densities for which the mean free path is equal to or greater than the interparticle separation, $l_c \gtrsim l_d$, as collision-triggered slowing down governs motility-induced phase separation.  This condition corresponds to average packing fractions below the close-collision threshold
$\phi_{\mathrm{cc}} = \pi/4$, which provides an upper bound slightly smaller than the hexagonal close-packing fraction
$\phi_{\mathrm{cp}} = \pi/(2\sqrt{3})$ . \\
Equations~(\ref{eq:position})-(\ref{eq:orientation}) are numerically integrated using a standard Milstein scheme suitable for stochastic differential equations with multiplicative noise. The time step is chosen in the range $\Delta t=10^{-4}$--$10^{-5}$ to ensure numerical stability and convergence. Our results are reported here on the long time limit, where transient effects due to initial conditions die out. We assumed that the simulation box length was much larger than period of the convection roll.  
\section{Results and Discussion}
We systematically analyze the structure and dynamics of an active suspension subjected to convective flow. In this section, we present and discuss the packing fraction distribution, density fluctuations, and effective diffusion coefficients, $D$, within the system. We explore the parameter space to obtain a comprehensive phase diagram that identifies the region in which the system exhibits a two-phase coexistence. Our key findings are presented on the basis of the simulation results reported in Figs.~\ref{Fig2}--\ref{Fig9}

\subsection*{Impact of Advection Strength on MIPS}
\begin{figure*}[t]
	\centering	\includegraphics[width=0.85\textwidth,height=0.85\textwidth]{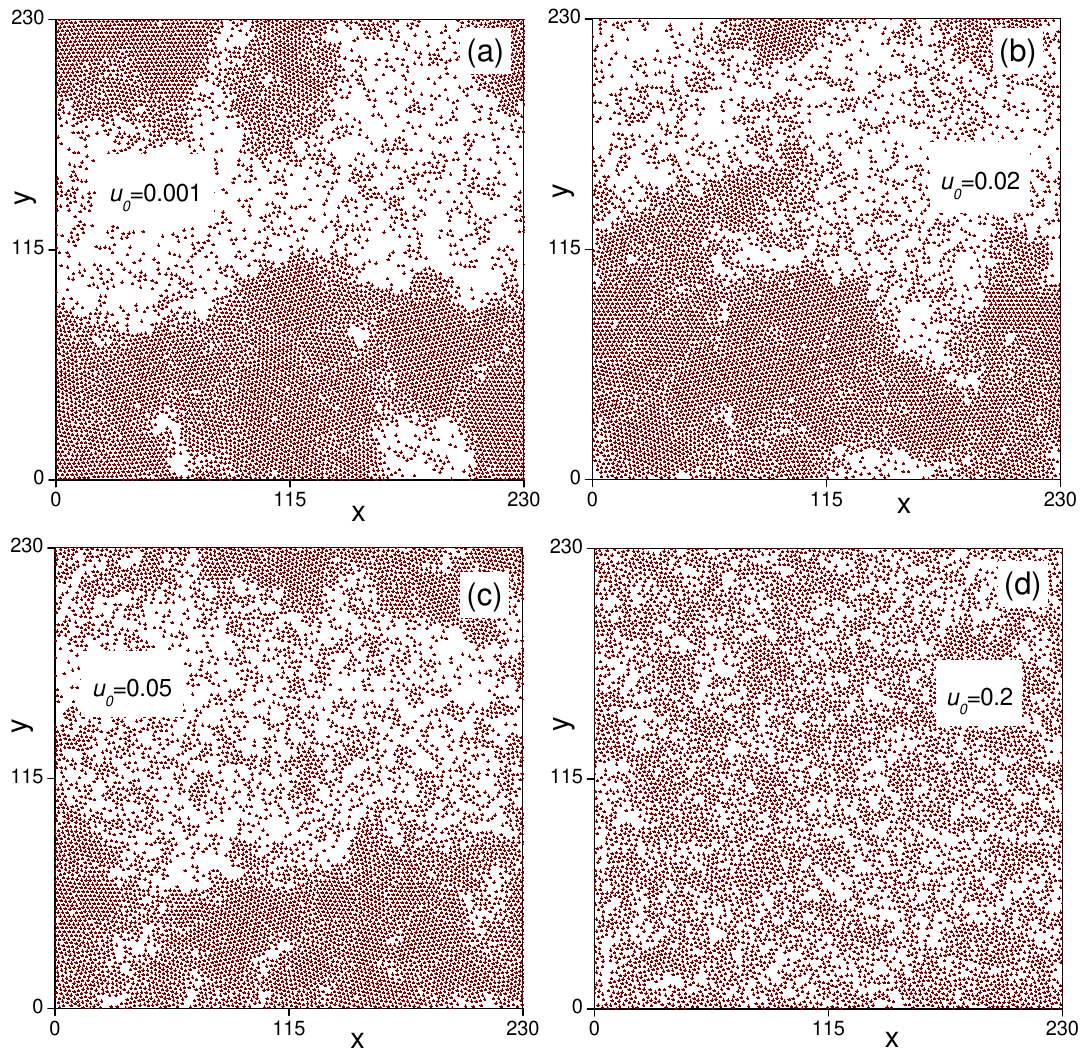}
    \caption{(a-d) Suspension of $N\sim8000$ active Janus particles with $\bar{\phi}=0.45$, $L=2\pi$ at various $u_0$. All the other simulation parameters are  
       $r_0=1.0$, $v_0=0.5$, $D_t=D_r=0.01$, $\epsilon =1.0$ and snapshots are shown at time $t=10^4$.}
    \label{Fig2}
\end{figure*}
The snapshots shown in Fig.~\ref{Fig2} demonstrate that advection strongly influences motility-induced phase separation. In Figs.~\ref{Fig2}(a–d), $v_0$ is held fixed at $\phi=0.45$. This value is chosen such that, $\phi\le\phi_{cc}$ and remains below, $\phi_{cp}$ thereby ensuring that any observed clustering is not driven by geometric constraints in the presence of flow. 

The advection speed, $u_0$ is systematically varied to probe its influence on the collective dynamics. Through the snapshots present in Fig.~\ref{Fig2}, we explore the regime from $0.002 \lesssim \chi \lesssim 0.4$, corresponding to weak-flow conditions, where particle motion is dominated by self-propulsion, to higher advection, which sufficiently suppresses MIPS.

In the weak-advection limit, the system clearly exhibits phase separation characterized by the coexistence of a dense and a dilute phase. In this regime, particle accumulation is primarily driven by self-propulsion. Following collisions, particles cannot readily pass through one another. Instead, their persistent motion causes them to continue exerting force on neighboring particles, leading to effective self-trapping within clusters. This trapping mechanism that maintains cluster stability remains largely unaffected at low advection. Consequently, large-scale phase coexistence is possible as particles remain confined in crowded areas for longer periods.

As the advection strength increases, flow-induced transport enhances the lateral redistribution of particles across convection rolls. Advection extracts particles from dense phases and destabilizes cluster interfaces, thereby reducing the spatial extent of the liquid-like phase. Consequently, clusters become progressively fragmented and short-lived, reflecting the growing competition between activity-driven accumulation and advective mixing.

For sufficiently strong advection $(u_0 \gtrsim 0.1$ for $v_0 = 0.5)$, flow suppresses the activity-induced local accumulation mechanism that is responsible for MIPS. As a result, dense clusters dissolve, and the system tends to a spatially homogeneous steady state . 
\subsection{Characterizing MIPS}

\subsubsection{Packing Fraction Distribution}

\begin{figure}[t]
    \centering
    \includegraphics[width=0.75\columnwidth]{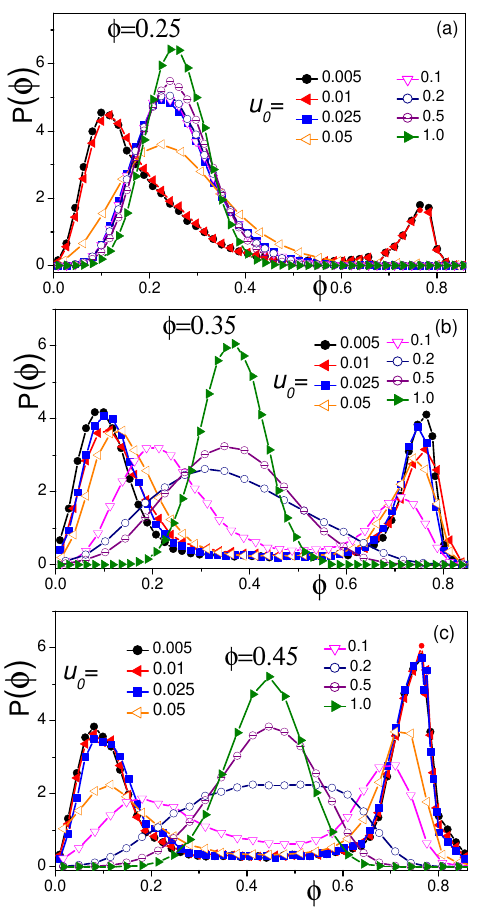}
    \caption{
    Probability distribution of the local packing fraction, $P(\phi)$, shown in panels (a)--(c) for systems with increasing average packing fraction. Panels (a), (b), and (c) correspond to $\phi = 0.25$, $0.35$, and $0.45$, respectively, for $v_0=1.0$. All other parameters are the same as in Fig~\ref{Fig2}.
    }
    \label{Fig3}
\end{figure}

To quantitatively characterize the suppression of MIPS by the imposed flow field, we analyze the local packing fraction distribution $P(\phi)$, for various $u_0$. The distribution allows us to track the transition from a phase-separated to a homogeneous state and precisely compute densities of the coexisting dilute and dense phases. Fig~\ref{Fig3}(a–c) shows $P(\phi)$ for systems with progressively increasing packing fraction. In the weak-advection regime, the distribution exhibits a distinct bimodal structure. The two well-separated peaks correspond to coexisting dilute (gas-like) and dense (liquid-like) phases.  This bimodality is a characteristic feature of MIPS capturing the contrasting local densities.

As $u_0$ increases, the bimodal distribution collapses into a single unimodal peak centered around the global average packing fraction $\phi$. The disappearance of bimodality indicates a complete breakdown of macroscopic phase separation and the formation of a  stable homogeneous state. In a completely homogeneous phase, the distribution is narrow and sharply centered around $\phi$, indicating spatial uniformity.

At lower global packing fractions $(\phi \ll \phi_{cc}$, see Fig~\ref{Fig3}a), phase separation is intrinsically less robust due to lower frequency of collision-induced crowding. Consequently, the crossover from bimodal to unimodal behavior occurs at significantly lower $u_0$ values compared to denser suspension. 

\begin{figure}
	\centering	\includegraphics[width=0.4\textwidth,height=0.3\textwidth]{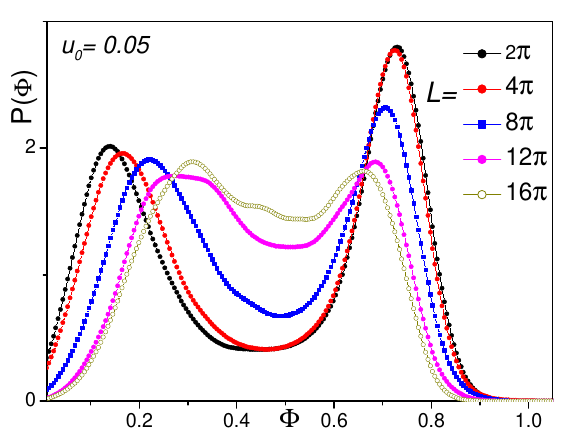}
    \caption{Packing fraction distribution $P(\phi)$ for different spatial periodicities, $L$ at fixed flow speed $u_0 = 0.05$, $v_0=1.0$. For smaller $L$, the distribution exhibits a pronounced bimodal structure, indicating strong phase separation into dense and dilute regions. As $L$ increases, the two peaks progressively move closer and the distribution becomes less distinctly bimodal, reflecting a weakening of MIPS.
 }
    \label{Fig4}
\end{figure}

In Fig~\ref{Fig4}, we examine the impact of spatial periodicity $L$ on the local packing fraction distribution $P(\phi)$ at a fixed advection speed. For smaller roll widths $(L=2\pi)$, the distribution displays sharp bimodal peaks, indicating strong phase separation. As $L$ increases, the coexisting peaks approach one another, signalling a gradual weakening of MIPS. An increase in $L$ results in a decrease in roll vorticity $\Omega_\text{ad}=2\pi u_0/L$, although the size of the convection cells increases simultaneously. 

As the vortical flow weakens, self-propulsion more effectively drives the particles. As a result, particle exchange between dense and dilute phases is enhanced. For $L=16\pi$, we observe a slight oscillation in $P(\phi)$, which might be associated with a periodic flow pattern. The spatial periodic structure becomes clearly evident with the strengthening of the advection velocity (see the supplementary movie).

This increased large-scale convective transport becomes more effective than self-trapping, thereby suppressing MIPS. 

\begin{figure}
	\centering	\includegraphics[width=0.4\textwidth,height=0.5\textwidth]{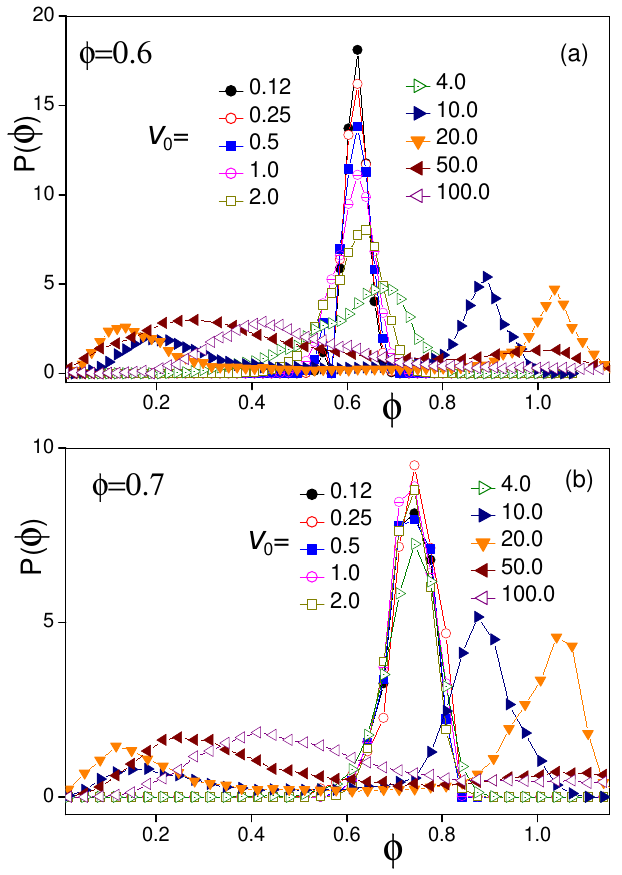}
    \caption{Probability distribution of the local packing fraction, $P(\phi)$, shown for systems with two different average packing fractions: (a) $\phi = 0.6$ and (b) $\phi = 0.7$. In each panel, the different curves illustrate the impact of varying the self-propulsion velocity, $v_0$, at a fixed advection strength $u_0 = 1.0$. All other parameters are identical to those in Fig.~\ref{Fig2}.}
    \label{Fig5}
\end{figure}

Next we investigate the effect of varying $v_0$ at a fixed $u_0$ for $\phi \lesssim \phi_{cc} $, we explore a broad range of $v_0$, as illustrated in Fig~\ref{Fig5}.  

At low $v_0$, the distribution remains unimodal with a peak near the mean packing fraction. In this regime, activity is insufficient to generate the collision-induced slowing necessary for phase separation, and the system remains homogeneous. As $v_0$ increases to intermediate values, the distribution develops a pronounced bimodal structure, signaling the coexistence of dense and dilute regions.   

However, upon further increasing $v_0$ the bimodality weakens, and the peaks shift toward the mean packing fraction. Extremely high propulsion leads to increased overlap within dense regions. This effectively softens excluded-volume interactions, enabling particles to escape dense regions more efficiently, thereby reducing cluster stability and weakening phase coexistence.  

Thus, the system exhibits reentrant behavior, wherein phase separation emerges at intermediate activity but is suppressed at sufficiently high self-propulsion, leading to a restoration of the homogeneous state. \\
\subsubsection*{Binodal Densities as a Function of Global Packing Fraction}
\begin{figure}[t]
    \centering \includegraphics[width=0.8\linewidth,height=0.8\textheight,keepaspectratio]{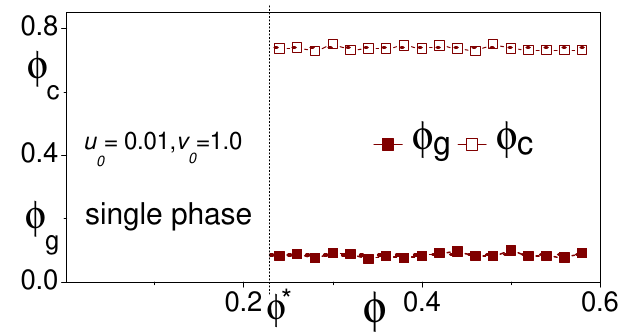}
    \caption{Binodal densities of the gas phase ($\phi_g$,filled squares) and cluster phase ($\phi_c$,open squares) as a function of the mean packing fraction $\phi$ at a fixed $u_0=0.01$ and $v_0=1.0$ and other parameters same as Fig.~\ref{Fig2}.The vertical dashed line marks the MIPS threshold $\phi^*$.  }
    \label{Fig6}
\end{figure}

To further characterize the phase-separated regime, we examine the behavior of the coexisting gas and cluster phase densities as a function of $\phi$, while keeping all other control parameters fixed. Fig~\ref{Fig6} shows the variation of the binodal densities as a function of $\phi$ in the range $\phi_{cp}>\phi>\phi^*$, where $\phi^*$ is the threshold packing fraction for MIPS. The coexistence densities $\phi_g$ and $\phi_c$ are extracted from the packing-fraction distribution $P(\phi)$, where the lower $\phi$ peak corresponds to the dilute (gas) phase and the higher $\phi$ peak corresponds to the dense (cluster) phase. 
A key observation is that both  $\phi_g$ and $\phi_c$ remain essentially constant for $\phi>\phi^*$, resulting in nearly constant horizontal lines. This indicates that, once phase separation sets in, the densities of the coexisting phases are independent of $\phi$. 
This behavior is physically characteristic of phase coexistence. When additional particles are introduced into the system beyond the MIPS threshold, the local densities of the individual phases do not change. Instead, their relative volume fractions adjust, allowing the dense cluster phase to grow at the expense of the dilute gas phase.

\subsubsection{Density Fluctuation}
To analyze density fluctuations, the simulation domain was coarse-grained by partitioning it into square compartments of linear size $l$, for each sub-box, the local particle number $n_i$ was computed and converted into a coarse-grained density as,
$\rho_i = n_i/V$, where $V=l^2$ is the area of each compartment. The mean density is given by, $\langle \rho_i \rangle = \langle n_i \rangle/V$
The variance of the coarse-grained density was evaluated as, $\langle (\delta \rho_i(l)^2 \rangle$ was then evaluated over all sub-boxes. By systematically varying the box size $l$, we obtained the scaling behavior of density fluctuations as a function of length scale.
\begin{equation}
\langle (\delta \rho_i)^2 \rangle
= \langle \rho_i^2 \rangle - \langle \rho_i \rangle^2 .
\end{equation}
Thus the density fluctuation can be expressed as,
\begin{equation}
\langle (\delta \rho_i)^2 \rangle
= \left\langle \left(\frac{n_i}{V}\right)^2 \right\rangle
- \left(\frac{\langle n_i \rangle}{V}\right)^2
= \frac{\langle n_i^2 \rangle - \langle n_i \rangle^2}{V^2}.
\label{eq:rho_var}
\end{equation}
This analysis was carried out for $\phi = 0.25, 0.35,$ and $0.45$ while varying $u_0$ over a broad range $(0.005 \leq u_0 \leq 1.0)$. \\
In a homogeneous system, using the microscopic definition of density and particle distribution functions~\cite{HansenMcDonald}, the variance of particle number in a subvolume can be expressed in terms of two-point correlations. For systems with finite correlation length, this leads to normal number fluctuations,
\begin{equation}
\langle n_i^2 \rangle - \langle n_i \rangle^2 \sim V .
\label{eq:number_fluc}
\end{equation}
Therefore we obtain,
\begin{equation}
\langle (\delta \rho_i)^2 \rangle \sim \frac{V}{V^2} = V^{-1} \sim l^{-2}.
\end{equation}

\begin{figure}[!t]
	\centering	
    \includegraphics[width=0.82\columnwidth]{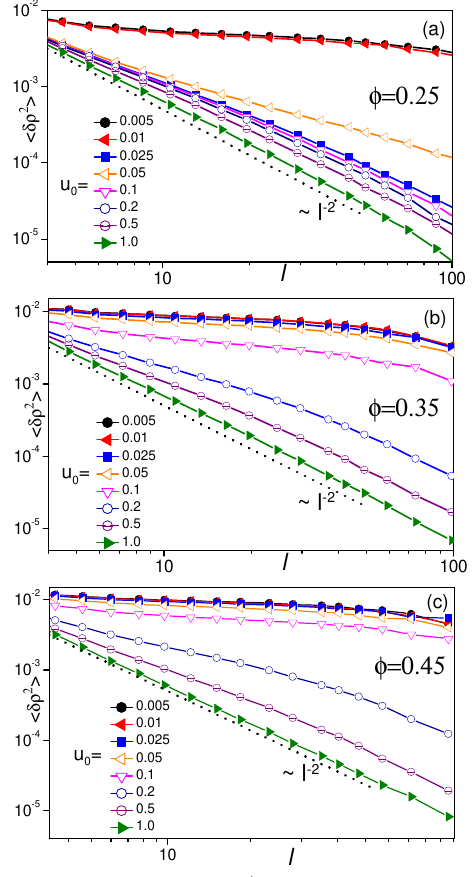} 
    \caption{
Density fluctuation plots, $\langle\delta\rho^2(l)\rangle$, shown in panels (a)–(c) for systems with increasing average packing fraction ($\phi$). Panels (a), (b), and (c) correspond to $\phi = 0.25$, $0.35$, and $0.45$, respectively. All other simulation parameters are identical to those used in Fig.~1. In panel (a), increasing the advection speed from $u_0 = 0.005$ to $0.01$ leads to a suppression of the decay from $l^{-2}$. For panels (b) and (c), a similar suppression is observed even at higher advection speeds, $u_0 = 0.05$--$0.1$. 
}
    \label{Fig7}
\end{figure}

Deviations from this $l^{-2}$ scaling are used to identify the onset of motility-induced phase separation. As the system approaches the MIPS regime, particle motility becomes increasingly sensitive to the local density, suppressing particle exchange between neighboring regions and enhancing spatial correlations.As a result, the decay of density fluctuations with $l$ becomes weaker than $l^{-2}$, and eventually saturates at large length scales, reflecting the emergence of macroscopic density inhomogeneities. \\ 
At low density ($\phi=0.25$, Fig.~\ref{Fig7}a), MIPS is observed only within a narrow range of $u_0$, as indicated by the suppression of the $l^{-2}$ decay. For higher values of $u_0$ $( u_0 > 0.01)$ the system becomes homogeneous, with fluctuations following near-equilibrium scaling. \\
At higher densities ($\phi=0.35, 0.45$, Fig.~\ref{Fig7}b,c) the range of $u_0$ exhibiting suppressed fluctuation decay broadens significantly. Deviations from $l^{-2}$ scaling persist up to $u_0 \approx 0.1$, which indicates an enhanced stability of the phase-separated state against advection. 
The crossover of density fluctuation from normal $l^{-2}$ scaling to suppressed decay of $l^{-1}$, followed by saturation at large length scales, provides a robust signature of phase separation.
\subsubsection{Diffusion}

\begin{figure}[t]
    \centering \includegraphics[width=0.8\linewidth,height=0.8\textheight,keepaspectratio]{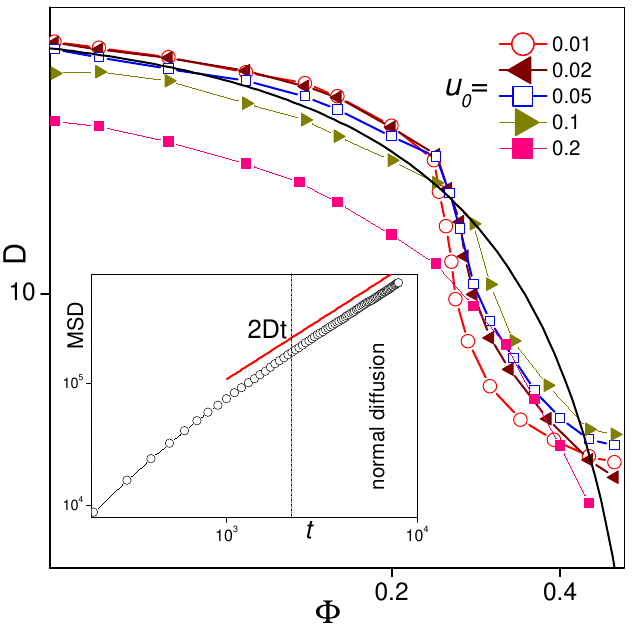}
    \caption{(color line) $D$ vs $\phi$ varying $u_0$ at $v_0=1.0$. All other simulation parameters are similar as Fig.~\ref{Fig1} Inset MSD vs $t$ plot depicting normal diffusion regime at $t \sim 10^4$ for $N\sim8000$ actve JP suspension. }
    \label{Fig8}
\end{figure}

 We analyze the effective diffusivity, $D$, as a function of $\phi$ to determine the MIPS threshold, $\phi^*$, in the presence of flow, as shown in Fig~\ref{Fig8}.

At high advection ($u_0 \geq 0.2$), the system remains homogeneous throughout a wide range of $\phi$ (Fig~\ref{Fig8}) and particle motion is spatially uniform. The corresponding diffusivity curve follows a smooth decrease with increasing $\phi$ due to enhanced collisions. This behaviour is well described by the quadratic fitting form \cite{stimuli-2}
\begin{equation}
D(\phi) = D_s \left(1 - \alpha \phi \right)^2,
\end{equation}
where $D_s = D_0 + v_0^2/2D_r$ represents the single-particle effective diffusivity in the dilute limit. Here, the term $v_0^2/(2D_r)$ captures the contribution of persistent self-propelled motion. The parameter $\alpha$ quantifies the reduction in mobility due to crowding and is related to the effective packing constraints in the system. This form reflects a gradual reduction of a particle's motion as the available free space decreases with increasing density.

In the presence of advection, deviations from this simple polynomial behaviour (solid black line)  emerge. At low advection, flow is insufficient to completely disrupt particle clustering. The system undergoes MIPS beyond a critical packing fraction $\phi^*$. For $\phi < \phi^*$ the system remains homogeneous, and the diffusivity continues to follow the smooth quadratic trend. However, a sharp reduction in $D$ is observed, at $\phi \sim \phi^*$ indicating the commencement of phase separation. For $\phi > \phi^*$ the diffusivity decreases rapidly and deviates considerably from the homogeneous-phase fit.

This sudden decrease in the diffusivity is due to the appearance of coexisting phases with different dynamical features. In the dense phase, particles get effectively trapped by persistent collisions and crowding, resulting in restricted motion. In contrast, particles in the dilute phase retain relatively high mobility. The overall behaviour $D(\phi)$ reflects contributions from both the homogeneous and phase-separated regimes.

Thus, the onset and extent of this phase separation are strongly influenced by the flow strength. Strong flows tend to homogenize the system, pushing $\phi^*$ to higher values or even suppressing phase separation. On the other hand at weaker advection, self propulsion dominates over advective transport, and the persistence induced trapping can promote phase separation at $\phi\sim\phi^*$.

Overall, the diffusivity as a function of $\phi$ offers an effective measure to detect the commencement of MIPS. The change from a smooth, quadratic decay in the homogeneous domain to a steeper decay in the phase-separated regime captures the change from uniform dynamics to heterogeneous, coexistence-driven dynamics.  

\subsubsection{Phase Diagram}
\begin{figure*}
    \centering \includegraphics[width=0.85\textwidth]{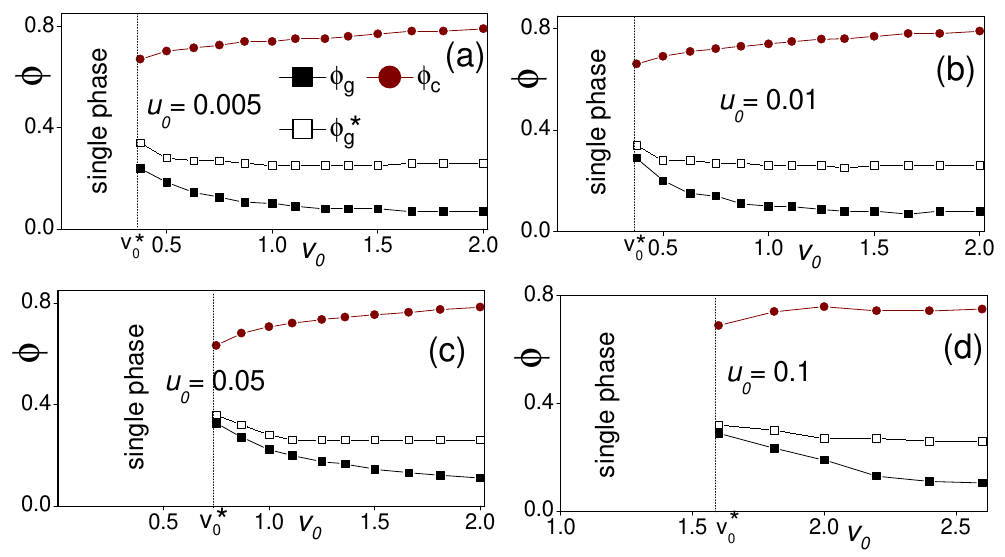}
    \caption{(a-d) Packing fraction,$\phi$ as a function of self-propulsion speed $v_0$, for various advection speed $u_0$ for $N=8000$, $D_0=D_r=0.01$. The filled symbols represent the binodal densities of the gas phase ($\phi_g$, black squares) and the cluster phase ($\phi_c$, red circles), while the open squares denote the gas spinodal $\phi_g^*$. The vertical dashed line marks the critical self-propulsion speed $v_0^*$, separating the homogeneous (single-phase) region from the phase-separated regime. For $v_0 < v_0^* $, the system remains homogeneous, whereas for $v_0 > v_0^* $, coexistence between dilute and dense phases emerges. The proximity of the cluster binodal and spinodal indicates a weakly distinguishable metastable region on the dense side.}
    \label{Fig9}
\end{figure*}

The phase behaviour of the system under advection is characterized through phase diagrams constructed in the parameter space of $\phi$ and $v_0$ for various $u_0$. These diagrams provide a systematic framework to understand the interplay between activity and flow on MIPS.

The spinodal line defines the boundary of stability of the homogeneous state. Below the spinodal line, the homogeneous state is stable and small density fluctuations decay with time. The spinodal line marks the onset of instability. Above the spinodal boundary, small density fluctuations grow spontaneously, eventually leading to cluster formation and phase separation.  

In Fig~\ref{Fig9}, the curve for \(\phi_g^*\) denotes the gas spinodal line, while \(\phi_g\) represents the corresponding binodal. The region bounded by these curves, \(\phi_g < \bar{\phi} < \phi_g^*\), is a metastable regime, where sufficiently large density fluctuations can induce phase separation. 

After identifying the boundaries between homogeneous and phase-separated states from the phase diagrams, we explore the influence of flow on the phase behavior. Comparison of the phase diagrams for different $u_0$ reveals a systematic shift of the critical self-propulsion velocity, \( v_{0}^* \) required for phase separation and the spatial extent of the coexistence region. 

As shown in Figs~\ref{Fig9}(a)–(d), increasing advection shifts $v_{0}^*$ towards higher values, indicating that stronger activity is required to destabilize the homogeneous state in the presence of flow. Consequently, the single-phase region expands with increasing $u_0$, and the coexistence region is shifted to larger values of $v_0$. These results demonstrate that advective transport effectively stabilizes the homogeneous phase and suppresses the onset of MIPS.

\section{Summary and Conclusion}

In this work, we have systematically investigated the role of convection rolls on the motility-induced phase separation (MIPS).          
The analysis of local packing fraction distributions and spatial density fluctuations clearly demonstrates the suppression of MIPS under imposed advection. In the absence of advection or within the weak-advection regime, the probability distribution $P(\phi)$ exhibits a distinct bimodal structure, signaling the coexistence of dense clusters and a dilute gas phase. For stronger advection, this bimodality disappears and is replaced by a narrow unimodal distribution centered at the mean global density, indicating the restoration of a structurally homogeneous configuration. This structural transition is accompanied by a systematic suppression of density fluctuations, evidenced by significant deviations from the characteristic $l^{-2}$ scaling observed in homogeneous systems. Near the MIPS threshold, reduced particle exchange between regions leads to suppressed fluctuations, leading to an eventual saturation at larger length scales. Furthermore, at $\phi \ll \phi_{cc}$, phase separation is inherently weaker, causing the transition from bimodal to unimodal behaviour to occur at significantly smaller values of $u_0$. Beyond the MIPS threshold, the binodal densities of the gas and cluster phases remain approximately constant, independent of the global packing fraction. This confirms that phase separation proceeds through a redistribution of particles between phases.
 
 By analyzing the diffusivity curves, we observed distinct dynamical changes marking each regime. In the homogeneous state, the diffusivity decreases smoothly with increasing $\phi$, following a quadratic scaling consistent with crowding-induced slowing down. In contrast, at lower advection strengths, a sharp drop in diffusivity at $\phi=\phi^*$ marks the onset of phase separation. This transition reflects the emergence of coexisting dense and dilute phases with markedly different dynamical properties.

Finally, phase diagrams constructed in the $(\phi, v_0)$ parameter space reveal that increasing advection strength systematically shifts $v_0^*$ to higher values, thereby expanding the stable single-phase region. The spinodal boundaries, which define the limits of linear instability, shrink progressively with increasing flow strength and can be suppressed entirely at sufficiently large $u_0$.   

Furthermore, at high self-propulsion velocity, we observe a reentrant behavior where phase separation weakens and eventually disappears. This is attributed to enhanced escape dynamics and a lower trapping efficiency, as particles can more readily leave dense regions. In systems with soft repulsive interactions, large propulsion further facilitates particle overlap and rearrangement, effectively softening excluded-volume constraints and destabilizing clusters.

Collectively, these observations demonstrate that the stability of MIPS under flow is governed by a competitive balance between activity, density, and advection transport. These insights provide an efficient strategy to dynamically control and tune collective non-equilibrium behavior in active synthetic and biological systems.
\section*{Supporting Information}
We provide supplementary movies illustrating the dynamic behavior for, $L=16\pi$ under strong advection for various global packing fraction $(\phi)$. The underlying convective flow becomes clearly distinguishable, with the  spatial distribution of particles directly mirroring the convective roll pattern. An additional movie, corresponding to the parameters in Fig.~\ref{Fig4}, shows continuous particle exchange between the dense and dilute phases. 

\section*{Acknowledgments}
PB thanks UGC India for the NET-JRF fellowship (NTA Ref No.: 211610009561).

\end{singlespace}
\end{document}